\documentclass[aps,singlecolumn,superscriptaddress,preprint,12pt]{revtex4-1}
\usepackage{amssymb}
\usepackage{amsmath}
\usepackage{graphicx}
\usepackage{epsfig}
\begin{document}
\title{Reciprocal and unidirectional scattering of parity-time symmetric structures}
\author{L. Jin}
\email{jinliang@nankai.edu.cn}
\affiliation{School of Physics, Nankai University, Tianjin 300071, China}
\author{X. Z. Zhang}
\affiliation{College of Physics and Materials Science, Tianjin Normal University, Tianjin
300387, China}
\author{G. Zhang}
\affiliation{School of Physics, Nankai University, Tianjin 300071, China}
\author{Z. Song}
\affiliation{School of Physics, Nankai University, Tianjin 300071, China}
\begin{abstract}
Parity-time ($\mathcal{PT}$) symmetry is of great interest.
The reciprocal and unidirectional features are intriguing besides the $\mathcal{PT}$ symmetry phase transition.
Recently, the reciprocal transmission, unidirectional reflectionless and
invisibility are intensively studied. Here, we show the reciprocal reflection/transmission
in $\mathcal{PT}$-symmetric system is closely related to the type of $\mathcal{PT}$ symmetry, that is, the axial (reflection) $\mathcal{PT}$ symmetry leads to reciprocal reflection (transmission).
The results are further elucidated by studying the scattering of rhombic ring form coupled resonators
with enclosed synthetic magnetic flux. The nonreciprocal phase shift induced by the magnetic flux and
gain/loss break the parity ($\mathcal{P}$) and time-reversal ($\mathcal{T}$) symmetry but keep the parity-time ($\mathcal{PT}$) symmetry.
The reciprocal reflection (transmission) and unidirectional transmission (reflection) are found in the axial (reflection) $\mathcal{PT}$-symmetric ring centre. The explorations of symmetry and asymmetry from $\mathcal{PT}$ symmetry may shed light on novel one-way optical devices and application of $\mathcal{PT}$ metamaterials.
\end{abstract}
\maketitle

\section*{Introduction}
Parity-time ($\mathcal{PT}$) symmetric quantum system may possess entirely
real spectrum although being non-Hermitian~\cite%
{Bender98,Bender02,AM02,MZnojil06,MZnojil08,Moiseyev,Jin09,Joglekar10,Gong10,Joglekar12,Gong13,SChen,Wang}%
. $\mathcal{PT}$ symmetric system is invariant under the combined $\mathcal{%
PT}$ operator in the presence of balanced gain and loss. In the past decade,
$\mathcal{PT}$-symmetric system has attracted tremendous interests as it
possesses unintuitive but intriguing implications.
Due to the similarity between the paraxial wave equation describing spatial light wave propagation
and the temporal Schr\"{o}dinger equation for quantum system, the complex refractive index distribution
satisfying $n^{\ast }(x)=n(-x)$ mimics $%
\mathcal{PT}$-symmetric potentials $V^{\ast }(x)=V(-x)$, $\mathcal{PT}$-symmetric systems
are proposed and realized in coupled optical waveguides through index guiding and a inclusion of
balanced gain and loss regions~\cite{Musslimani OL,Makris08,AGuo,CERuter}. A number of novel and non-trivial phenomena are found, such as power oscillation~%
\cite{CERuter}, coherent perfect absorbers~\cite{CPA,CPAlonghi,Hasan},
nonreciprocal light propagation~\cite{Regensburger} in coupled
waveguides, and recently the $\mathcal{PT}$-symmetric microcavity
lasing~\cite{Jing,Zhang,Hodaei} and gain induced large optical nonlinear~%
\cite{PengNP,Peng,Chang,Jing2015,ZhangJ2015PRB} in coupled resonators.

The spectral singularity~\cite%
{AliPRL2009,UnidirectionalSS,AliPRA2011,SSLonghi,SSAhmed,AliPRL2013,SSAli}
and invisibility~\cite%
{Lin,Feng,LonghiInvisibility10,LonghiInvisibility14,AliInvisibility13,AliInvisibility14,AliInvisibility15}
in $\mathcal{PT}$-symmetric system are hot topics, where reciprocal
transmission and unidirectional reflectionless in $\mathcal{PT}$-symmetric
metamaterial are intriguing features for novel optical devises. These devices are useful for
light transport, control and manipulation~\cite%
{Kottos2012R,Kottos2012,ReciprocalAhmed,ReciprocalAli}. The symmetric
scattering properties are usually attributed to certain internal symmetry of
a scattering centre. For instance, the parity ($\mathcal{P}$) symmetry, or
time-reversal ($\mathcal{T}$) symmetry of a scattering centre leads to
symmetric reflection and transmission~\cite{LXQ} ($\mathcal{T}$-symmetric
system without unequal tunnelling amplitude is Hermitian, otherwise, only
reciprocal reflection or transmission holds~\cite{PTscattering}). Here, we report reciprocal reflection, similar as reciprocal transmission, are both related to the $\mathcal{PT}$ symmetry of a scattering
centre: The axial (refection) $\mathcal{PT}$ symmetry, with respect to the
input and output channels, induces reciprocal reflection (transmission).
Recent efforts on photonic Aharonov-Bohm effect enable photons
behaving like electrons in magnetic field. Effective magnetic field for photons can be
introduced in coupled waveguides by bending the waveguides~\cite{ABcaging},
periodically modulating the refractive index~\cite{DynModuPRL}, and the photon-phonon interactions~\cite{PhotPhon}; or in coupled resonators by
magneto-optical effect~\cite{MOeffect}, dynamic modulation~\cite{DynModu},
and off-resonance coupling paths imbalance~\cite{ChongYD,Hafezi2014}.
In this work, we focus on the $\mathcal{PT}$-symmetric structure with balanced gain and loss
threading by synthetic magnetic flux, where photons feel a
nonreciprocal tunnelling phase between neighbour resonators. The nonreciprocal
tunnellings and balanced gain and loss break the $\mathcal{P}$ and $\mathcal{T}$
symmetry but keep the $\mathcal{PT}$ symmetry of the scattering centre. The axial
(reflection) $\mathcal{PT}$ symmetry will lead to reciprocal reflection
(transmission) and unidirectional transmission (reflection). Our findings
provide new insights of $\mathcal{PT}$ symmetry and the symmetric/asymmetric
scattering, which are instrumental for the applications of $\mathcal{PT}$%
-symmetric metamaterial for light transport and one-way optical devises.

\section*{Results}
\begin{figure*}[tb]
\includegraphics[ bb=0 0 515 310, width=16.8 cm, clip]{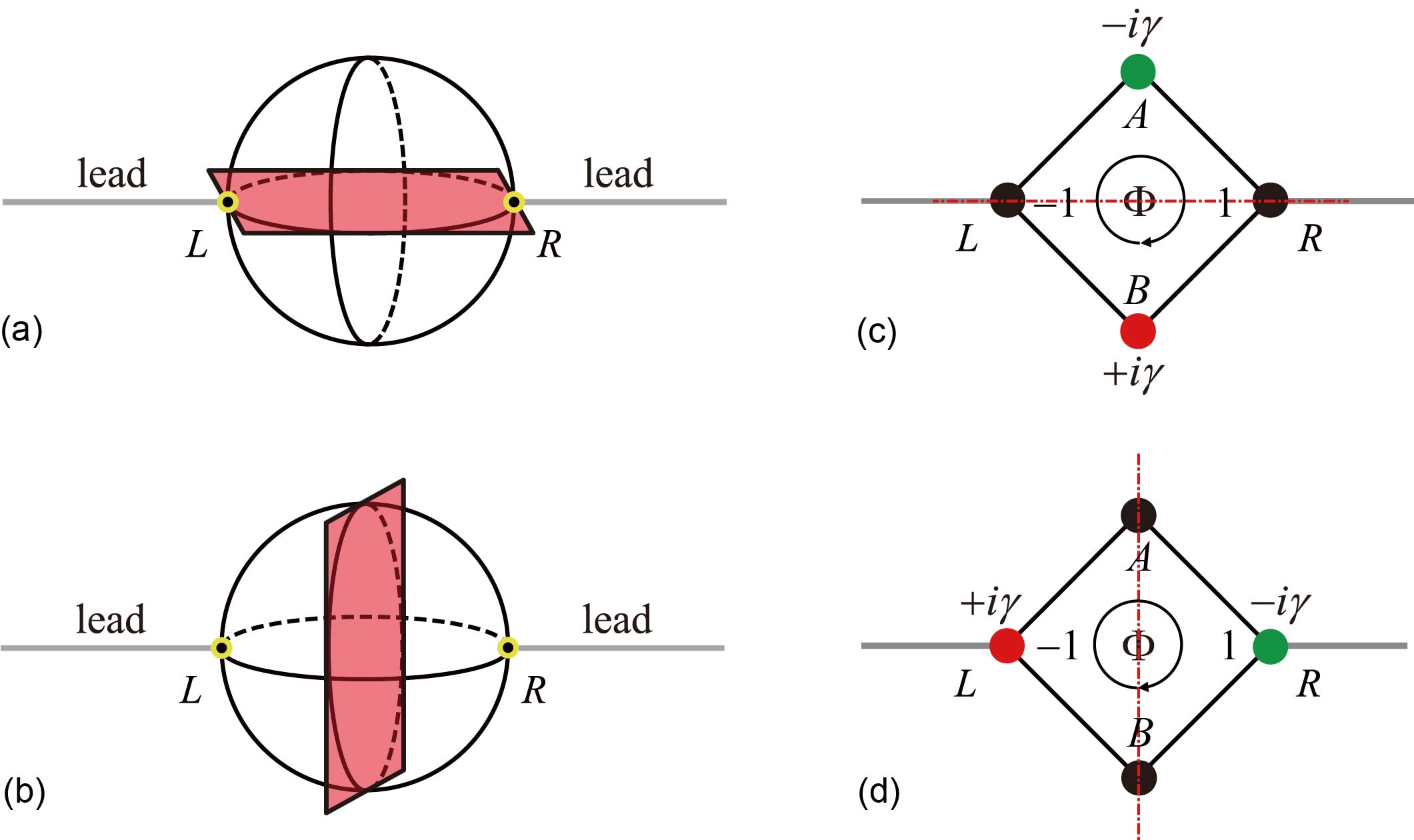}
\caption{\textbf{The type of $\mathcal{PT}$ symmetry.} Two semi-infinite
input and output leads (solid grey) are connected to the $\mathcal{PT}$%
-symmetric structures (black sphere) at sites $L$ and $R$ (yellow circle). (a) The
axial $\mathcal{PT}$ symmetry, defined as self-correspondence of $L,R $
under parity operation. (b) The reflection $\mathcal{PT}$ symmetry, defined
as reflection-correspondence of $L,R $ under parity operation.
The axial (c) and reflection (d) $\mathcal{PT}$-symmetric
rhombic ring configurations with enclosed magnetic flux $\Phi$ are schematically illustrated. The $\mathcal{PT}$ symmetry axes are in dash dotted red. The red (green) site represents the gain (loss).}
\label{fig1}
\end{figure*}

\textbf{Reciprocal and unidirectional scattering of $\mathcal{PT}$%
-symmetric structures.} The symmetric scattering properties of a $\mathcal{PT}
$-symmetric structure are closely related to the classification of $\mathcal{%
PT}$-symmetry. The parity operator $\mathcal{P}$ is the spatial reflection operator, $\mathcal{T}$ is the time-reversal
operator. In Fig.~\ref{fig1}a,b, we schematically show two types of $\mathcal{PT%
}$ symmetry. The Hamiltonian of the scattering centre is $\mathcal{PT}$%
-invariant, i.e., $\left( \mathcal{PT}\right) H_{\mathrm{c}}\left( \mathcal{%
PT}\right) ^{-1}=H_{\mathrm{c}}$. The input and output leads are connected to
the $\mathcal{PT}$-symmetric scattering centre at sites $L$ and $R$. If the
connection sites under the parity operation satisfies $\mathcal{P}L\mathcal{P%
}^{-1}=L$, $\mathcal{P}R\mathcal{P}^{-1}=R$, the system is called axial $%
\mathcal{PT}$ symmetric (Fig.~\ref{fig1}a). If the connection sites under
the parity operation satisfies $\mathcal{P}L\mathcal{P}^{-1}=R$, $\mathcal{P}%
R\mathcal{P}^{-1}=L$, the system is called reflection $\mathcal{PT}$
symmetric (Fig.~\ref{fig1}b). The red plane indicates the up-to-down (left-to-right) spatial reflection correspondence of axial (reflection) $%
\mathcal{PT}$ symmetry.

In order to address the reciprocal and unidirectional scattering behavior.
We study the reflection and transmission of a scattering centre for the left side and right side inputs, respectively. We denote the two scattering states as $%
\psi _{\mathrm{L}}^{k}$ and $\psi _{\mathrm{R}}^{k}$ for the input with wave vector $%
k $. The forward going and backward going waves are in form of $e^{\pm ikj}$%
. Combining with the reflection and transmission coefficient, we assume the
scattering state wave function (not at the spectral singularities, see
Methods) of left side input as,
\begin{equation}
\psi _{\mathrm{L}}^{k}\left( j\right) =\left\{
\begin{array}{c}
e^{ikj}+r_{\mathrm{L}}e^{-ikj},j<0 \\
t_{\mathrm{L}}e^{ikj},j>0%
\end{array}%
\right. .  \label{Psi_L}
\end{equation}%
where $r_{\mathrm{L}}$ and $t_{\mathrm{L}}$ represent the reflection and
transmission coefficients for the left side input with wave vector $k$.
Similarly, the wave function of right side input is in form of%
\begin{equation}
\psi _{\mathrm{R}}^{k}\left( j\right) =\left\{
\begin{array}{c}
t_{\mathrm{R}}e^{-ikj},j<0 \\
e^{-ikj}+r_{\mathrm{R}}e^{ikj},j>0%
\end{array}%
\right. .  \label{Psi_R}
\end{equation}%
where $r_{\mathrm{R}}$ and $t_{\mathrm{R}}$ represent the reflection and
transmission coefficients for the right side input with wave vector $k$.

\textbf{Reciprocal reflection under axial $\mathcal{PT}$\ symmetry.}
As shown in Fig.~\ref{fig1}a, this type of $\mathcal{PT}$-symmetric
scattering centres have connection sites under parity operation
satisfying $\mathcal{P}L\mathcal{P}^{-1}=L$, $\mathcal{P}R\mathcal{P}^{-1}=R$%
. In the axial $\mathcal{PT}$-symmetric configuration, the $\mathcal{PT}$
symmetry is defined as $\left( \mathcal{PT}\right) H_{\mathrm{L}%
}\left( \mathcal{PT}\right) ^{-1}=H_{\mathrm{L}}$, $\left( \mathcal{PT}%
\right) H_{\mathrm{R}}\left( \mathcal{PT}\right) ^{-1}=H_{\mathrm{R}}$ in the leads, and as $\left( \mathcal{PT}\right) H_{\mathrm{c}}\left( \mathcal{PT%
}\right) ^{-1}=H_{\mathrm{c}}$ in the centre. The whole scattering system is axial $%
\mathcal{PT}$-symmetric with respect to the leads. The axial $\mathcal{PT}$
symmetry results in symmetric relations on the scattering coefficients as (see Supplementary Information Note 1 for details),
\begin{eqnarray}
|r_{\mathrm{L}}|^{2}+t_{\mathrm{R}}t_{\mathrm{L}}^{\ast } &=&1,
\label{axialPT1} \\
|r_{\mathrm{R}}|^{2}+t_{\mathrm{L}}t_{\mathrm{R}}^{\ast } &=&1.
\label{axialPT2}
\end{eqnarray}%
From equations~(\ref{axialPT1},\ref{axialPT2}), we notice the reflection
probabilities for the left and right side inputs are the same, i.e.,%
\begin{equation}
|r_{\mathrm{L}}|^{2}=|r_{\mathrm{R}}|^{2}.
\end{equation}%
In other words, the axial $\mathcal{PT}$ symmetry leads to the reciprocal
reflection. Notice that we have reciprocal reflection $|r_{\mathrm{L}%
}|^{2}=|r_{\mathrm{R}}|^{2}=1$ at any transmission zero $t_{\mathrm{L,R}}=0$
, where one-way pass through is possible. Furthermore,considering the waves with vectors $k$
and $-k$, the reflection and transmission coefficients further satisfy $r_{%
\mathrm{L}}(-k)=r_{\mathrm{L}}^{\ast }( k) $, $t_{\mathrm{L}}(-k)=t_{\mathrm{%
L}}^{\ast }(k)$, $r_{\mathrm{R}}(-k)=r_{\mathrm{R}}^{\ast }(k) $, and $t_{%
\mathrm{R}}(-k)=t_{\mathrm{R}}^{\ast }(k)$.

\textbf{Reciprocal transmission under reflection $\mathcal{PT}$
symmetry.} As shown in Fig.~\ref{fig1}b, this type of $\mathcal{PT}$%
-symmetric scattering centres have connection sites under parity
operation satisfying $\mathcal{P}L\mathcal{P}^{-1}=R$, $\mathcal{P}R\mathcal{%
P}^{-1}=L$. In the reflection $\mathcal{PT}$-symmetric configuration, the $%
\mathcal{PT}$ symmetry is defined as $\left( \mathcal{PT}%
\right) H_{\mathrm{L}}\left( \mathcal{PT}\right) ^{-1}=H_{\mathrm{R}}$, $%
\left( \mathcal{PT}\right) H_{\mathrm{R}}\left( \mathcal{PT}\right) ^{-1}=H_{%
\mathrm{L}}$ in the leads, and as $\left( \mathcal{PT}\right) H_{\mathrm{c}%
}\left( \mathcal{PT}\right) ^{-1}=H_{\mathrm{c}}$ in the centre. The whole scattering
system is reflection $\mathcal{PT}$-symmetric with respect to the leads. The
reflection $\mathcal{PT}$ symmetry results in symmetric relations on the
scattering coefficients as (see Supplementary Information Note 2 for
details),
\begin{eqnarray}
|t_{\mathrm{L}}|^{2}+r_{\mathrm{R}}r_{\mathrm{L}}^{\ast } &=&1,
\label{reflectionPT1} \\
|t_{\mathrm{R}}|^{2}+r_{\mathrm{L}}r_{\mathrm{R}}^{\ast } &=&1.
\label{reflectionPT2}
\end{eqnarray}%
From equations~(\ref{reflectionPT1}, \ref{reflectionPT2}), we notice the
transmission probabilities for the left side and right side inputs are the same,
i.e.,
\begin{equation}
|t_{\mathrm{L}}|^{2}=|t_{\mathrm{R}}|^{2}.
\end{equation}%
This indicates the reflection $\mathcal{PT}$ symmetry leads to the reciprocal transmission, as observed in Bragg gratings and other $%
\mathcal{PT}$-symmetric structures~\cite%
{ReciprocalAhmed,ReciprocalAli,Kottos2012}. Notice that we have reciprocal
transmission $|t_{\mathrm{L}}|^{2}=|t_{\mathrm{R}}|^{2}=1$ at any reflection
zero $r_{\mathrm{L,R}}=0$, where unidirectional reflectionless is possible~%
\cite%
{Lin,Feng,LonghiInvisibility10,LonghiInvisibility14,AliInvisibility13,AliInvisibility14,AliInvisibility15}%
. Furthermore, considering the waves with vectors $k$ and $-k$, the reflection and
transmission coefficients further satisfy $r_{\mathrm{L}}(-k)=r_{\mathrm{R}%
}^{\ast }(k) $, $t_{\mathrm{L}}(-k)=t_{\mathrm{R}}^{\ast}(k)$, $r_{\mathrm{R}%
}(-k)=r_{\mathrm{L}}^{\ast }(k) $, and $t_{\mathrm{R}}(-k)=t_{\mathrm{L}%
}^{\ast }(k)$.

We show that in the present of $\mathcal{PT}$ symmetry,
the reciprocal reflection or transmission in a scattering centre is protected when the axial or reflection
$\mathcal{PT}$ symmetry holds even though the $\mathcal{P}$ and $\mathcal{T}$
symmetry are absent. Moreover, $\mathcal{PT}$
symmetry structure may exhibit unidirectional scattering behavior.

\textbf{$\mathcal{PT}$-symmetric rhombic ring structures.} We use
a rhombic ring structure (Fig.~\ref{fig1}c,d) to elucidate the results. The scattering centre
encloses with an effective magnetic flux $\Phi$, photons moving along the rhombic
ring structure in clockwise (counterclockwise) direction will acquire an additional direction-dependent phase factor $e^{+i\Phi }$ ($e^{-i\Phi }$), thus photons tunnelling is nonreciprocal except when $\Phi
=n\pi$, $n\in\mathbb{Z}$. This is an
effective photon Aharonov-Bohm effect creating by synthetic magnetic field~\cite%
{ABcaging,DynModuPRL,PhotPhon,MOeffect,DynModu,ChongYD,Hafezi2014}. The phase factor $e^{\pm i\Phi }$ is an analytical function of $\Phi $ with
period of $2\pi $, it is sufficient to understand the influence of magnetic
flux on the scattering by studying $\Phi $ in the region $[0,2\pi )$. To realize a synthetic magnetic field, two ring resonators are
coupled through an auxiliary off-resonant ring resonator. The auxiliary
resonator introduces optical paths imbalance when coupling to two primary resonators, the auxiliary resonator can be effectively reduced and create a coupling phase factor between two primary resonators. The coupled resonators under synthesized magnetic field is described by a magnetic tight-binding Hamiltonian~\cite%
{ChongYD,Hafezi2014},
\begin{equation}
H_{0}=-e^{i\phi }(a_{1}^{\dagger }a_{A}+a_{A}^{\dagger
}a_{-1}+a_{-1}^{\dagger }a_{B}+a_{B}^{\dagger }a_{1})+\mathrm{h.c.,}
\label{H_0}
\end{equation}%
where $\phi =\Phi /4$ is a nonreciprocal phase shift induced
by the magnetic flux in the tunnelling constant. In Fig.~\ref{fig1}c, the Hamiltonian of the scattering centre is $H_{%
\mathrm{c}}=H_{0}-i\gamma a_{A}^{\dagger }a_{A}+i\gamma a_{B}^{\dagger
}a_{B} $, where $\gamma $ is the gain/loss rate. The balanced gain and loss are the origin
of the non-Hermiticity realized in the optical systems~\cite{Musslimani
OL,Makris08,AGuo,CERuter,PengNP,Zhang,Hodaei,Peng,Chang}. The configuration is axial $\mathcal{PT}$-symmetric
with the parity operator acting on the rhombic ring sites defined as $\mathcal{P}%
\left( -1\right) \mathcal{P}^{-1}=-1$, $\mathcal{P}A\mathcal{P}^{-1}=B$, $%
\mathcal{P}1\mathcal{P}^{-1}=1$,$\mathcal{P}B\mathcal{P}^{-1}=A$. In Fig.~%
\ref{fig1}d, the Hamiltonian of the
scattering centre is $H_{\mathrm{c}}=H_{0}+i\gamma a_{-1}^{\dagger
}a_{-1}-i\gamma a_{1}^{\dagger }a_{1}$, the parity operator $\mathcal{P}$ is
defined as $\mathcal{P}\left( -1\right) \mathcal{P}^{-1}=1$, $\mathcal{P}A%
\mathcal{P}^{-1}=A$, $\mathcal{P}1\mathcal{P}^{-1}=-1$, $\mathcal{P}B%
\mathcal{P}^{-1}=B$, and the configuration is reflection $\mathcal{PT}$%
-symmetric. In the system, the magnetic flux is inverted meanwhile the gain and loss are switched under the $\mathcal{P}$ or $\mathcal{T}$ operation. However, the system is invariant under the combined $\mathcal{PT}$ operator, i.e., the presence of non-trivial magnetic flux as well as balanced gain and loss both break the $\mathcal{P}$ and $\mathcal{T}$ symmetry but keep the $\mathcal{PT}$ symmetry of the system.

The scattering centre is actually a two-arm Aharonov-Bohm interferometer. Light wave propagates through two pathways ($A$ and $B$) between the connection sites $-1$, $1$ and interfere with each other. The interference generates the output which varies as the enclosed magnetic flux. The effective magnetic field is gauge invariant and the magnetic flux acts globally, thus the reflection and transmission are not affected by the nonreciprocal phase distribution in the tunnellings for fixed magnetic flux. In the following, we discuss the reflection and transmission of the $\mathcal{PT}$-symmetric rhombic ring structures in details.

\begin{figure*}[tb]
\includegraphics[ bb=0 0 525 315, width=16.8 cm, clip]{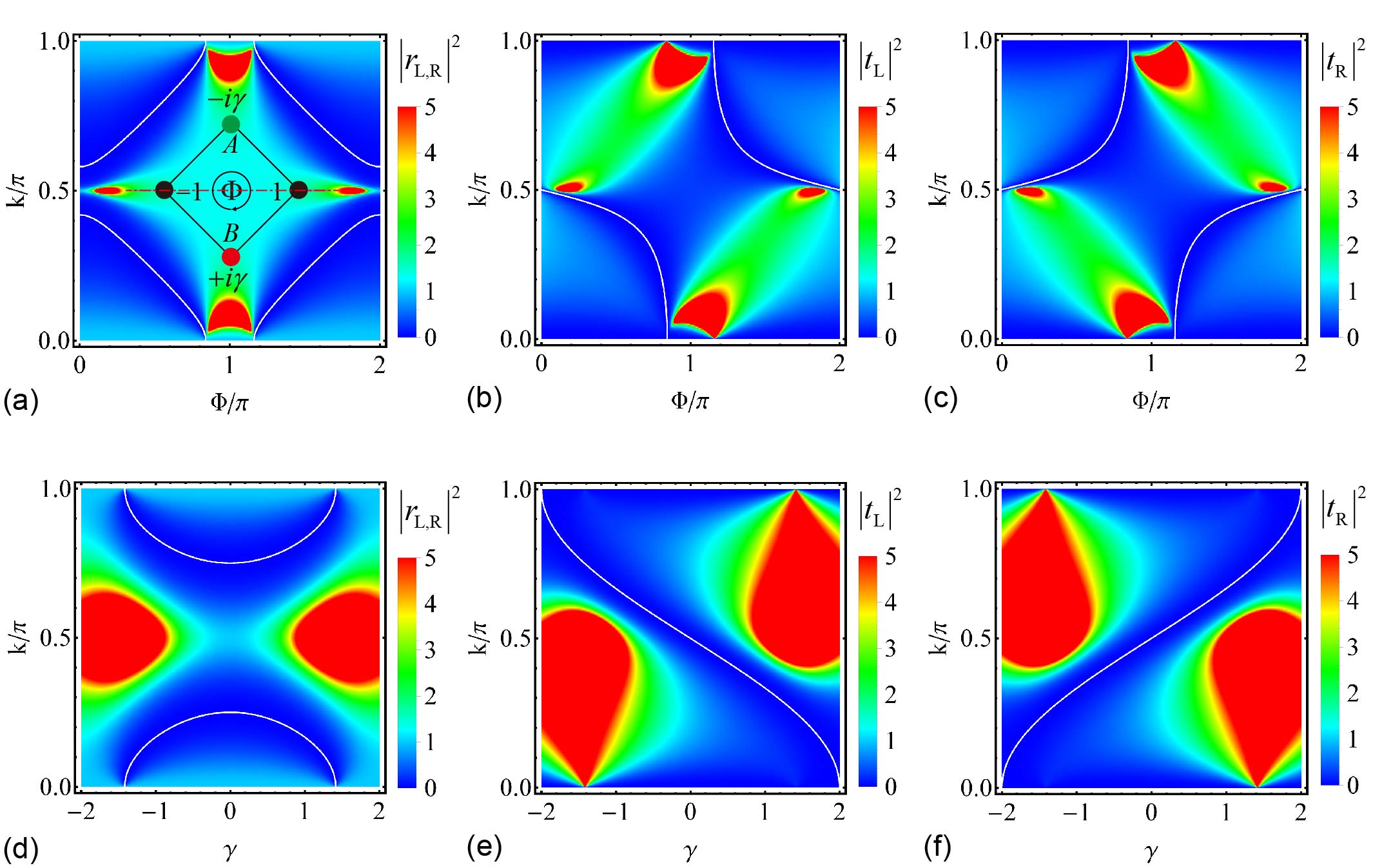}
\caption{\textbf{Reciprocal reflection and unidirectional transmission under
axial $\mathcal{PT}$ symmetry.} (a-c) The reflection and transmission
probabilities $|r_{\mathrm{L,R}}|^{2}$, $|t_{\mathrm{L}}|^{2}$, $|t_{\mathrm{%
R}}|^{2}$ at $\protect\gamma =1/2$ as functions of $\Phi $ and $k$. (d-f)
The reflection and transmission probabilities $|r_{\mathrm{L,R}}|^{2}$, $|t_{%
\mathrm{L}}|^{2}$, $|t_{\mathrm{R}}|^{2}$ at $\Phi =\protect\pi /2$ as
functions of $\protect\gamma $ and $k$. The insert in (a) schematically
illustrates the axial $\mathcal{PT}$-symmetric rhombic ring structure. The white curves are the reflection and transmission zeros. At $k=\protect\pi /2$ and $%
\Phi =0,2\pi$, the reflections in (a) are $1$, the transmissions in (b,c) are $0$.}
\label{fig2}
\end{figure*}

The reflection and transmission coefficients for the axial $\mathcal{PT}$-symmetric rhombic ring structure (Fig.~\ref{fig1}c) are calculated from the Schr\"{o}dinger equations (see Methods), yielding
\begin{eqnarray}
r_{\mathrm{L}} &=&r_{\mathrm{R}}=\frac{\gamma ^{2}+4[\sin ^{2}k-\cos
^{2}(\Phi /2)]}{4e^{2ik}\cos ^{2}(\Phi /2)+4\sin ^{2}k-\gamma ^{2}},
\label{axialPTrLR} \\
t_{\mathrm{L}} &=&\frac{4i\sin k[2\cos k\cos (\Phi /2)-\gamma \sin (\Phi /2)]%
}{4e^{2ik}\cos ^{2}(\Phi /2)+4\sin ^{2}k-\gamma ^{2}},  \label{axialPTtL} \\
t_{\mathrm{R}} &=&\frac{4i\sin k[2\cos k\cos (\Phi /2)+\gamma \sin (\Phi /2)]%
}{4e^{2ik}\cos ^{2}(\Phi /2)+4\sin ^{2}k-\gamma ^{2}}.  \label{axialPTtR}
\end{eqnarray}%
The reflection and transmission probabilities are functions of the magnetic flux $%
\Phi $, gain/loss rate $\gamma $, and wave vector $k$. They satisfy $%
\left\vert r_{\mathrm{L}}\left( \Phi ,\gamma ,k\right) \right\vert
^{2}=\left\vert r_{\mathrm{R}}\left( \Phi ,\gamma ,k\right) \right\vert ^{2}$
(see Fig.~\ref{fig2}a,d), $\left\vert t_{\mathrm{L}}\left( \Phi ,\gamma
,k\right) \right\vert ^{2}=\left\vert t_{\mathrm{R}}\left( -\Phi ,\gamma
,k\right) \right\vert ^{2}$ (see Fig.~\ref{fig2}b,c), and $\left\vert t_{%
\mathrm{L}}\left( \Phi ,\gamma ,k\right) \right\vert ^{2}=\left\vert t_{%
\mathrm{R}}\left( \Phi ,-\gamma ,k\right) \right\vert ^{2}$ (see Fig.~\ref%
{fig2}e,f). In the absence of non-trivial magnetic flux $\Phi $, or gain/loss $%
\gamma $, the system is $\mathcal{P}$-symmetric (reflection-$\mathcal{P}$%
-symmetric for $\Phi $ absence, i.e., left to right by mirror imaging;
inversion-$\mathcal{P}$-symmetric for $\gamma $ absence, i.e., left to right
by $180{^{\circ }}$ rotation), the reflection and transmission are both
reciprocal. The non-trivial magnetic flux $\Phi $ together with balanced gain and loss $\gamma $ break the $\mathcal{P}$ symmetry. The symmetric
transmission in the $\mathcal{PT}$-symmetric system at $k\neq \pi /2$ is
broken, i.e., the transmission is unidirectional at $k\neq \pi /2$. Moreover, the axial $\mathcal{PT}$ symmetry protects the symmetric reflection, therefore, the reflection is reciprocal but the transmission is unidirectional. The
white curves in Fig. \ref{fig2} show the reflection/transmission zeros. At $%
\gamma =\pm 2\cos k\cot (\Phi /2)\neq 0$, we have $t_{\mathrm{L}}=0$ or $t_{%
\mathrm{R}}=0 $ with total reflection $\left\vert r_{\mathrm{L,R}%
}\right\vert ^{2}=1$. This indicates that we only have a non-zero transmission for the right side or left side input, thus the axial $\mathcal{PT}$-symmetric rhombic ring structure allows one-way pass through.

Photons circle in the scattering centre either in a clockwise direction or in a counterclockwise direction,
we schematically illustrate the two pathways in the Supplementary (Fig.~\ref{figS1}). The phase difference between two pathways affects
the interference in the scattering centre, thus the transmission varies as the effective magnetic flux
induced phase difference. The phase difference between clockwise direction and counterclockwise direction of transmission pathways is $\Phi$ for
the left side input (Supplementary, see Fig.~\ref{figS1}c) or $-\Phi$ for the right side input (Supplementary, see Fig.~\ref{figS1}d).
The transmission pathways are not equivalent in the presence of gain/loss,
the interference of phase difference being $\Phi$ is different from the interference of phase difference being $-\Phi$.
Therefore, the unidirectional transmission is enabled in the presence of nonreciprocal tunneling phase
factor ($e^{+i\Phi}\neq e^{-i\Phi}$) attributed to non-trivial magnetic flux ($\Phi \neq n\pi$, $n\in\mathbb{Z}$).
\begin{figure*}[tb]
\includegraphics[ bb=0 0 525 155, width=16.8 cm, clip]{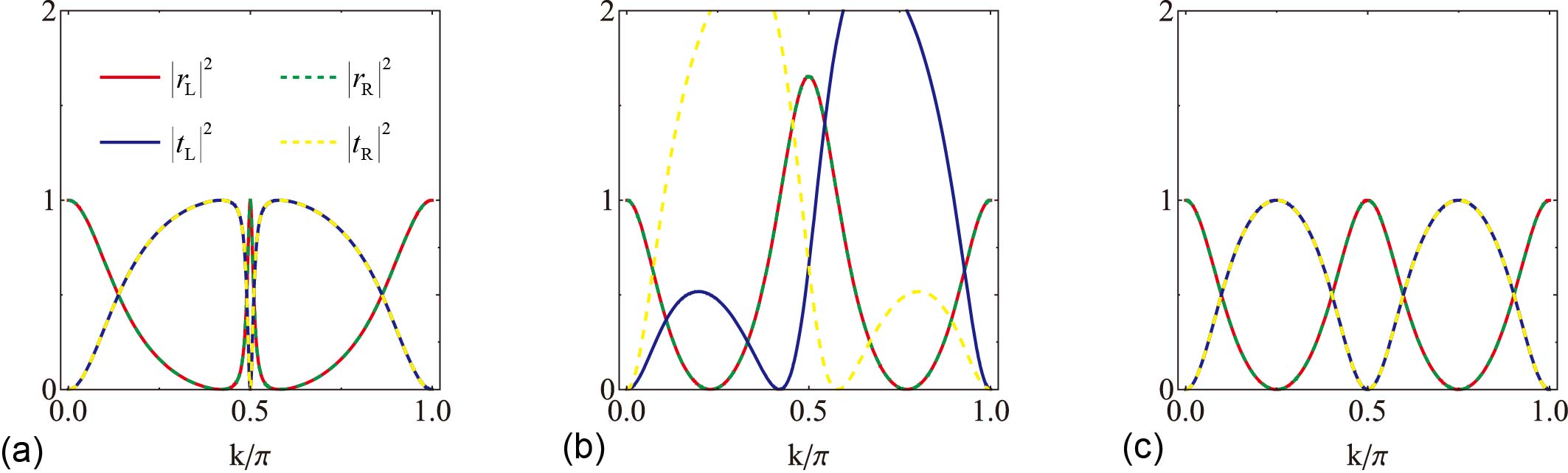}
\caption{\textbf{Symmetric reflection under axial $\mathcal{PT}$ symmetry.}
(a) $\protect\gamma =1/2$, $\Phi =0$, (b) $\protect\gamma =1/2$, $\Phi =%
\protect\pi /2$, (c) $\protect\gamma =0$, $\Phi =\protect\pi /2$.}
\label{fig3}
\end{figure*}

In Fig.~\ref{fig3}, we plot the reflection and transmission probabilities for an axial $\mathcal{PT}$-symmetric rhombic ring structure (Fig.~\ref{fig1}c) at several set
parameters. Figure~\ref{fig3}a is for a system with balanced gain and loss
in the absence of magnetic flux, i.e., $\gamma =1/2$, $\Phi =0$. The gain/loss, being non-Hermitian, plays the role of on-site potentials and is the origin of unidirectional behavior. However, the presence of balanced gain and loss alone does not ensure unidirectional scattering. We notice the reflection and transmission in Fig.~\ref{fig3}a are both reciprocal. The scattering is unitary even though the system is non-Hermitian (the balanced gain and
loss of this rhombic ring structure $i\gamma (A^{\dagger}A-B^{\dagger}B)$
can be reduced to an anti-Hermitian interaction $i\gamma (A^{\prime \dagger
}B^{\prime }+B^{\prime \dagger }A^{\prime })$ by composing $A^{\prime
},B^{\prime }=(B\pm A)/\sqrt{2}$, and the non-Hermiticity of the scattering
centre only arises from the anti-Hermitian interaction between $A^{\prime
},B^{\prime }$, which is proved to have unitary scattering~\cite%
{UnitaryScatJL}). By introducing magnetic flux to the system, the $\mathcal{P}$ and $\mathcal{T}$ symmetry is destroyed but the $\mathcal{PT}$ symmetry holds. The interference between light waves from the loss arm and the gain arm generates unidirectional transmission for non-trivial magnetic flux. Figure~\ref{fig3}b is for a system in the presence of non-trivial magnetic flux, i.e., $\gamma =1/2$, $\Phi
=\pi /2$. The unidirectional transmission zero happens at $k=\arccos (\pm
1/4)\pi $, i.e., at $k\approx 0.420\pi $, $\left\vert t_{\mathrm{L}%
}\right\vert ^{2}=0$, $\left\vert t_{\mathrm{R}}\right\vert ^{2}\approx
1.369 $; at $k\approx 0.580\pi $, $\left\vert t_{\mathrm{L}}\right\vert
^{2}\approx 1.369$, $\left\vert t_{\mathrm{R}}\right\vert ^{2}=0$, which
indicates a one-way pass through. Figure~\ref{fig3}c is for a Hermitian
scattering centre in the presence of non-trivial magnetic flux, i.e., $\gamma =0$, $\Phi =\pi /2$, we have Hermitian scattering without unidirectional behavior.

In the rhombic ring structure under axial $\mathcal{PT}$ symmetry (Fig. \ref{fig1}c), the reflection and transmission coefficients $r_{\mathrm{L}}$, $r_{\mathrm{R}}$, $t_{\mathrm{L}}$, $t_{\mathrm{R}}$ diverge at the spectral singularities~\cite%
{AliPRL2009}. When $k=\pi /2$, we have the reflection and transmission
coefficients $r_{\mathrm{L}}=r_{\mathrm{R}}=[4\sin ^{2}(\Phi /2)+\gamma
^{2}]/[4\sin ^{2}(\Phi /2)-\gamma ^{2}]$ and $t_{\mathrm{L}}=-t_{\mathrm{R}%
}=-4i\gamma \sin (\Phi /2)/[4\sin ^{2}(\Phi /2)-\gamma ^{2}]$. We notice the spectral
singularities are at $\gamma =\pm 2\sin \left( \Phi /2\right) \neq 0$. When
$\Phi =\pi $, we have the reflection and transmission coefficients $r_{%
\mathrm{L}}=r_{\mathrm{R}}=(4\sin ^{2}k+\gamma ^{2})/(4\sin ^{2}k-\gamma
^{2})$ and $t_{\mathrm{L}}=-t_{\mathrm{R}}=-4i\gamma \sin k/(4\sin
^{2}k-\gamma ^{2})$. The spectral singularities are at $\gamma =\pm 2\sin
k\neq 0$. At the spectral singularities, the scattering states are in form of a
self-sustained emission $f_{j<0}^{k}=e^{-ikj}$, $f_{j>0}^{k}=\mp ie^{ikj}$
and a reflectionless absorption $f_{j<0}^{k}=e^{ikj}$, $f_{j>0}^{k}=\pm
ie^{-ikj}$ (see Method)~\cite{ZXZPRA2013}. The transfer matrix
of the scattering centre is $M_{11}=M_{22}=0$, $M_{12}=\mp i$, $M_{21}=\pm i$
with matrix-element $M_{22}$ vanishes.

\begin{figure*}[tb]
\includegraphics[ bb=0 0 525 315, width=16.8 cm, clip]{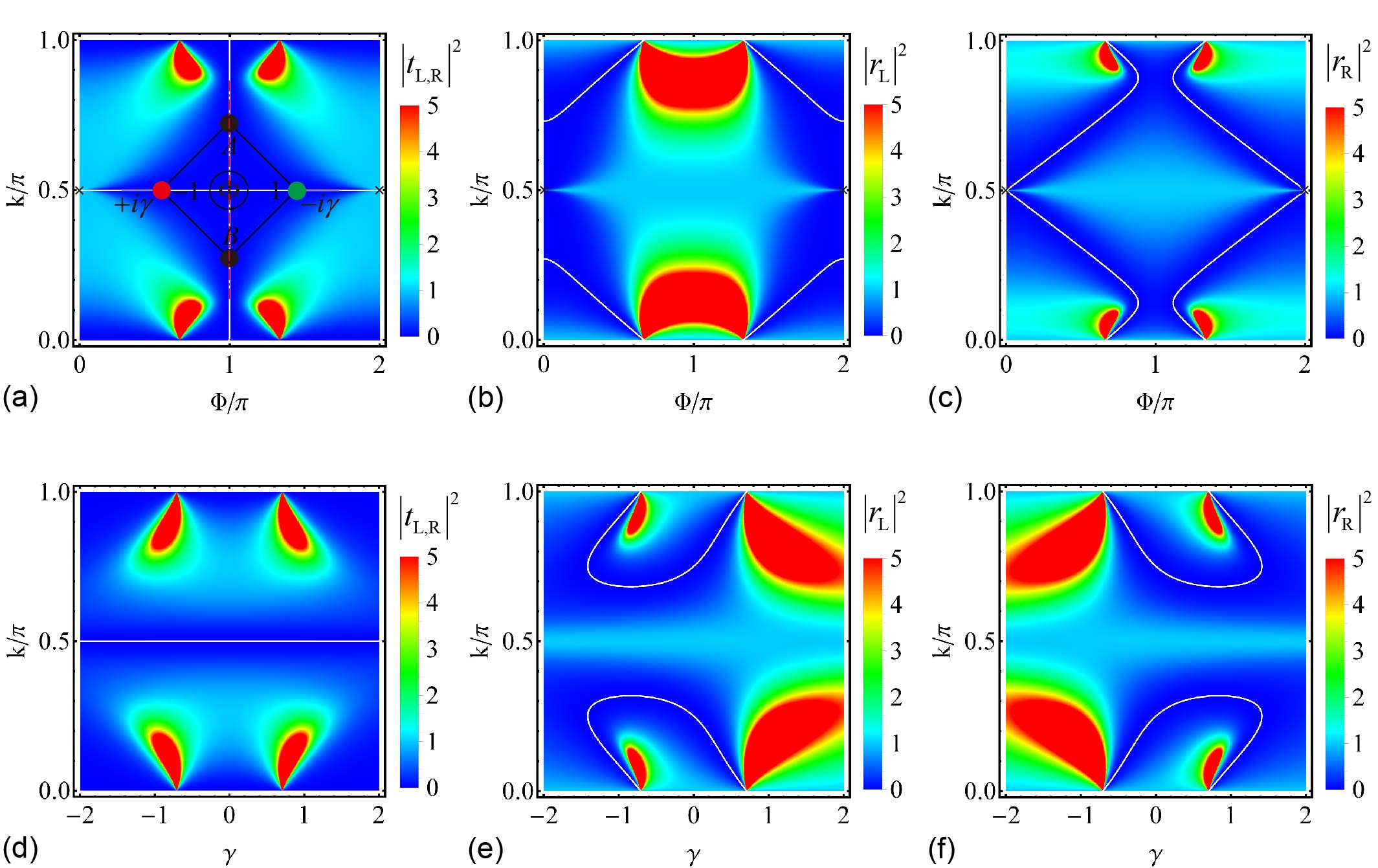}
\caption{\textbf{Reciprocal transmission and unidirectional reflection under
reflection $\mathcal{PT}$ symmetry.} (a-c) The reflection and transmission
probabilities $|t_{\mathrm{L,R}}|^{2}$, $|r_{\mathrm{L}}|^{2}$, $|r_{\mathrm{%
R}}|^{2}$ at $\protect\gamma =1/2$ as functions of $\Phi $ and $k$. (d-f)
The reflection and transmission probabilities $|t_{\mathrm{L,R}}|^{2}$, $|r_{%
\mathrm{L}}|^{2}$, $|r_{\mathrm{R}}|^{2}$ at $\Phi =\protect\pi /2$ as
functions of $\protect\gamma $ and $k$. The insert in (a) schematically
illustrates the reflection $\mathcal{PT}$-symmetric rhombic ring structure. The white curves are the reflection and
transmission zeros. At $k=\protect\pi /2$ and $\Phi =0,2\pi$, the transmissions in (a) are $1$, the reflections in (b,c) are $0$ as marked by black crosses.}
\label{fig4}
\end{figure*}

Now, we turn to discuss the rhombic ring structure under reflection $\mathcal{PT}$ symmetry.
The configuration is shown in Fig.~\ref{fig1}d.
In Supplementary, Figure~\ref{figS2} schematically illustrates the pathways of photons.
The connection sites are linked by two same pathways. In the presence of magnetic flux $\Phi$, photons travelling from left lead to right lead in clockwise direction and
counterclockwise direction acquire additional phases $+\Phi/2$ and $-\Phi/2$
in the two pathways (Supplementary, see Fig.~\ref{figS2}c), respectively.
The situation is unchanged for photons travelling inversely from right lead to left lead (Supplementary, see Fig.~\ref{figS2}d).
Equivalently, the upper and lower pathways are undistinguishable. Therefore, only relative phase difference
$\Phi$ matters (affecting the transmission coefficient) and the transmission is directionless.
The reflection and transmission coefficients are calculated as (see Methods)
\begin{eqnarray}
r_{\mathrm{L}} &=&\frac{(\gamma ^{2}+2\gamma \sin k-1)\cos ^{2}k+\sin
^{2}(\Phi /2)}{\sin ^{2}k-e^{2ik}[\gamma ^{2}\cos ^{2}k-\cos ^{2}(\Phi /2)]},
\label{reflectionPTrL} \\
r_{\mathrm{R}} &=&\frac{(\gamma ^{2}-2\gamma \sin k-1)\cos ^{2}k+\sin
^{2}(\Phi /2)}{\sin ^{2}k-e^{2ik}[\gamma ^{2}\cos ^{2}k-\cos ^{2}(\Phi /2)]},
\label{reflectionPTrR} \\
t_{\mathrm{L}} &=&t_{\mathrm{R}}=\frac{i\sin \left( 2k\right) \cos (\Phi /2)%
}{\sin ^{2}k-e^{2ik}[\gamma ^{2}\cos ^{2}k-\cos ^{2}(\Phi /2)]}.
\label{reflectionPTtLR}
\end{eqnarray}%
The reflection and transmission coefficients are functions of the magnetic flux $%
\Phi $, gain/loss rate $\gamma $, and wave vector $k$. They satisfy $%
\left\vert t_{\mathrm{L}}\left( \Phi ,\gamma ,k\right) \right\vert
^{2}=\left\vert t_{\mathrm{R}}\left( \Phi ,\gamma ,k\right) \right\vert ^{2}$
and $\left\vert r_{\mathrm{L}}\left( \Phi ,\gamma ,k\right) \right\vert
^{2}=\left\vert r_{\mathrm{R}}\left( \Phi ,-\gamma ,k\right) \right\vert
^{2} $. Figure~\ref{fig4} implies a reciprocal transmission (Fig.~\ref{fig4}%
a,d) and unidirectional reflection (Fig.~\ref{fig4}b,c,e,f). In this
configuration, the scattering with both reflection and transmission being
reciprocal happens in the absence of gain and loss ($\gamma =0$), that is when the
system is $\mathcal{P}$-symmetric. In the presence of gain and loss ($\gamma
\neq 0$), the reflection probability is unidirectional, but the reflection $%
\mathcal{PT}$ symmetry protects the reciprocal transmission. Due to the
presence of gain and loss, the probability of the total reflection and
transmission after scattering is not unitary, being balanced gain and loss
rate dependent. The white curves in Fig. \ref{fig4} show the reflection and transmission zeros. At $k=\protect\pi /2$ and $\Phi =0$, $| r_{\mathrm{L,R}}| ^{2}=0$, $|t_{\mathrm{L,R}}| ^{2}=1$. At the reflection zeros, $|r_{\mathrm{L(R)}}|^{2}=0$, $|
r_{\mathrm{R(L)}}|^{2}\neq 0$, and $| t_{\mathrm{L,R}}| ^{2}=1$, the system
exhibits unidirectional reflectionless with reciprocal transmission.

\begin{figure*}[tb]
\includegraphics[ bb=0 0 525 155, width=16.8 cm, clip]{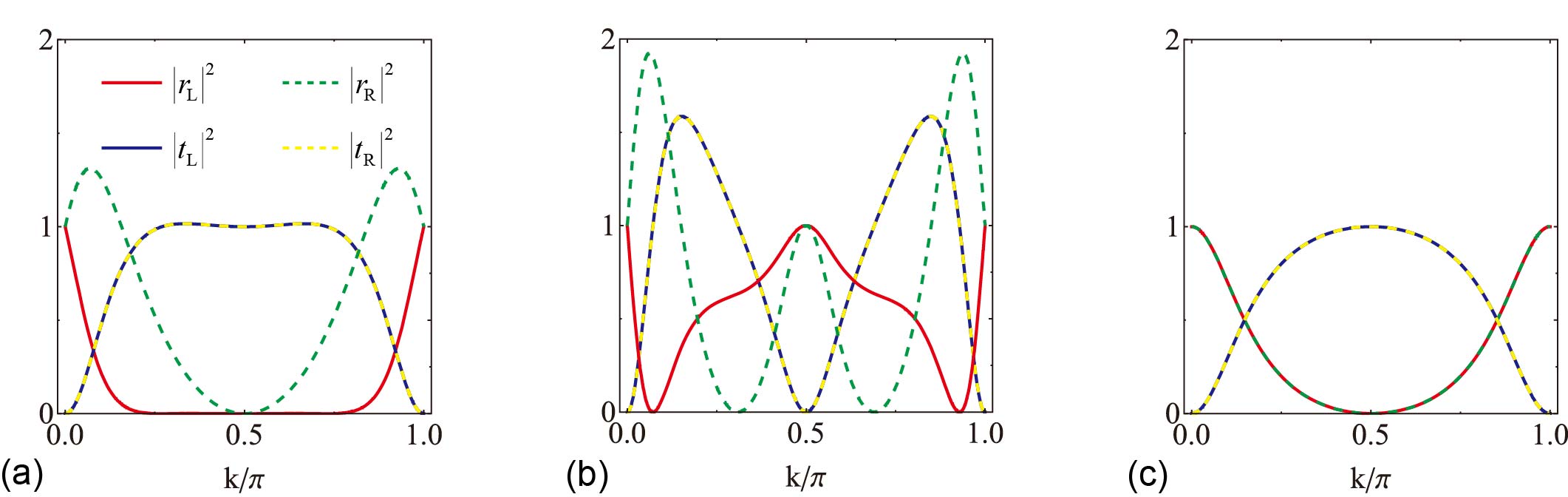}
\caption{\textbf{Symmetric transmission under reflection $\mathcal{PT}$
symmetry.} (a) $\protect\gamma =1/2$, $\Phi =0$, (b) $\protect\gamma =1/2$, $%
\Phi =\protect\pi /2$, (c) $\protect\gamma =0,\Phi =0$.}
\label{fig5}
\end{figure*}

In Fig.~\ref{fig5}, we plot the reflection and transmission probabilities for a reflection $\mathcal{PT}$-symmetric rhombic ring structure (Fig.~\ref{fig1}d) at several set parameters. Figure~\ref{fig5}a,b are for balanced gain and loss
rate $\gamma =1/2$ with two different magnetic flux $\Phi =0$
and $\pi /2$, respectively. The reciprocal transmission and unidirectional reflection are
clearly seen. In Fig.~\ref{fig5}a, the unidirectional reflectionless happens
at $k\approx 0.27\pi ,0.73\pi $, $|r_{\mathrm{L}}|^{2}=0,|r_{\mathrm{R}%
}|^{2}=0.437$, $|t_{\mathrm{L}}|^{2}=|t_{\mathrm{R}}|^{2}=1$. In Fig.~\ref{fig5}b, the unidirectional reflectionless
happens at $k\approx 0.072\pi ,0.928\pi $, $|r_{\mathrm{L}}|=0,|r_{\mathrm{R}%
}|=1.899$, $|t_{\mathrm{L}}|^{2}=|t_{\mathrm{R}}|^{2}=1$; or at $k\approx
0.310\pi ,0.690\pi $, $|r_{\mathrm{L}}|^{2}=0.634,|r_{\mathrm{R}}|^{2}=0$, $%
|t_{\mathrm{L}}|^{2}=|t_{\mathrm{R}}|^{2}=1$. In the absence of gain and
loss $\gamma =0$, the reflection and axial $\mathcal{PT}$-symmetric rhombic
ring configurations reduce to an identical system. In Fig.~\ref{fig5}c, we
plots the reflection and transmission probabilities of a scattering centre
in the absence of both gain and loss and magnetic flux, i.e., $\gamma =0$, $\Phi =0$, we observe Hermitian
scattering behavior of reciprocal reflection and transmission similar as $%
\gamma =0$, $\Phi =\pi /2$ shown in Fig.~\ref{fig3}c. Notice that no spectral
singularity emerges in the scattering of reflection $\mathcal{PT}$-symmetric rhombic ring
system. The system with $\Phi =0$ leads to input with wave vector $k=\pi /2$ both
sides invisible that $r_{\mathrm{L}}=r_{\mathrm{R}}=0$, $t_{\mathrm{L}}=t_{%
\mathrm{R}}=1$ (black crosses in Fig.~\ref{fig4}a-c); The system with $\Phi \neq 0$ leads to input with vector $%
k=\pi /2$ both sides opaque that $r_{\mathrm{L}}=r_{\mathrm{R}}=1$, $t_{%
\mathrm{L}}=t_{\mathrm{R}}=0$. For the input with wave vector $k=\pi /2$, the
scattering behavior is very sensitive to the magnetic flux.

\section*{Conclusion}

We investigate the reciprocal and unidirectional scattering of $\mathcal{PT}$%
-symmetric structures. We show an insightful understanding of the symmetric
scattering behavior, that is associated with the type of $\mathcal{PT}$
symmetry, defined as the $\mathcal{PT}$ symmetry of the
connection sites on the $\mathcal{PT}$-symmetric structures.
We find that the axial (reflection) $\mathcal{PT}$ symmetry leads to reciprocal
reflection (transmission). The transmission (reflection) is unidirectional affected
by the magnetic flux and gain/loss, this is because the magnetic flux induced nonreciprocal phase and the gain/loss break the $\mathcal{P}$ or $\mathcal{T}$ symmetry of the scattering centre.
The results are further elucidated using a $\mathcal{PT}$-symmetric rhombic ring structure with enclosed effective magnetic flux describing by tight-binding model. The physical realization of such scattering centre is possible in optical
systems such as coupled waveguides array and coupled resonators. Notice that
our conclusions are also applicable to the system with nonreciprocal
tunnelling being unequal tunnelling amplitude~\cite{ZXZAN2013}. We believe
our findings may shed light on coherent light transport and would be useful
for applications of quantum devices with inherent symmetry, in particular,
for novel unidirectional optical device designs that not limited to optical diodes using synthetic $\mathcal{PT}$-symmetric metamaterial.

\section*{Methods}

\textbf{Schr\"{o}dinger equations. } The input and output leads are
described by two semi-infinite tight-binding chain. The left lead is $H_{%
\mathrm{L}}=-\sum_{j=-\infty }^{-1}(a_{j-1}^{\dagger }a_{j}+a_{j}^{\dagger
}a_{j-1})$, the right lead is $H_{\mathrm{R}}=-\sum_{j=1}^{\infty
}(a_{j}^{\dagger }a_{j+1}+a_{j+1}^{\dagger }a_{j})$, where $a_{j}^{\dagger }$
($a_{j}$) is the creation (annihilation) operator of the site $j$, the
tunnelling between sites is uniform and set unity. The Hamiltonian of the
scattering system is $H=H_{\mathrm{L}}+H_{\mathrm{c}}+H_{\mathrm{R}}$. The
eigenstate of the scattering system is set $| \psi _{\mathrm{L,R}%
}^{k}\rangle =\sum_{j=-\infty }^{+\infty }f_{j}^{k}a_{j}^{\dagger
}\left\vert \mathrm{vac}\right\rangle +f_{A}^{k}a_{A}^{\dagger }\left\vert
\mathrm{vac}\right\rangle +f_{B}^{k}a_{B}^{\dagger }\left\vert \mathrm{vac}%
\right\rangle $.

For the axial $\mathcal{PT}$-symmetric configuration shown in Fig.~\ref{fig1}c, the Hamiltonian of the scattering system is $H_{\mathrm{c}%
}=H_{0}-i\gamma a_{A}^{\dagger }a_{A}+i\gamma a_{B}^{\dagger }a_{B} $. The
Schr\"{o}dinger equations $H| \psi _{\mathrm{L,R}}^{k}\rangle =E_{k}|\psi _{%
\mathrm{L,R}}^{k}\rangle $ on the scattering centre yield four independent
equations%
\begin{eqnarray}
-f_{-2}^{k}-e^{-i\phi }f_{A}^{k}-e^{i\phi }f_{B}^{k} &=&E_{k}f_{-1}^{k},
\label{L1} \\
-f_{2}^{k}-e^{i\phi }f_{A}^{k}-e^{-i\phi }f_{B}^{k} &=&E_{k}f_{1}^{k},
\label{L2} \\
-e^{i\phi }f_{-1}^{k}-e^{-i\phi }f_{1}^{k} &=&\left( E_{k}+i\gamma \right)
f_{A}^{k},  \label{L3} \\
-e^{-i\phi }f_{-1}^{k}-e^{i\phi }f_{1}^{k} &=&\left( E_{k}-i\gamma \right)
f_{B}^{k},  \label{L4}
\end{eqnarray}

For the reflection $\mathcal{PT}$-symmetric configuration shown in Fig.~\ref{fig1}d, the Hamiltonian of the scattering centre is $H_{\mathrm{c}%
}=H_{0}+i\gamma a_{-1}^{\dagger }a_{-1}-i\gamma a_{1}^{\dagger }a_{1}$.
Correspondingly, four independent
equations from the Schr\"{o}dinger equations $H| \psi _{\mathrm{L,R}%
}^{k}\rangle =E_{k}| \psi _{\mathrm{L,R}}^{k}\rangle $ on the scattering
centre are in form of%
\begin{eqnarray}
-f_{-2}^{k}-e^{-i\phi }f_{A}^{k}-e^{i\phi }f_{B}^{k} &=&\left( E_{k}-i\gamma
\right) f_{-1}^{k},  \label{R1} \\
-f_{2}^{k}-e^{i\phi }f_{A}^{k}-e^{-i\phi }f_{B}^{k} &=&\left( E_{k}+i\gamma
\right) f_{1}^{k},  \label{R2} \\
-e^{i\phi }f_{-1}^{k}-e^{-i\phi }f_{1}^{k} &=&E_{k}f_{A}^{k},  \label{R3} \\
-e^{-i\phi }f_{-1}^{k}-e^{i\phi }f_{1}^{k} &=&E_{k}f_{B}^{k},  \label{R4}
\end{eqnarray}%
where $\phi =\Phi /4$. The Schr\"{o}dinger equations on the leads give the energy $E_{k}=-2\cos k$
for the input with wave vector $k$. Notice that $k=\pi/2$ in the reflection $\mathcal{PT}$-symmetric
rhombic ring with $\Phi=2n\pi$ ($n\in\mathbb{Z}$) results in $f_{-1}=-f_{1}$ and the transmissions are 1. Otherwise, $k=\pi/2$ in system with $\Phi\neq 2n\pi$ ($n\in\mathbb{Z}$) leads to $f_{-1}=f_{1}=0$ and the transmissions are 0.\newline

\noindent \textbf{Reflection and transmission coefficients. }To calculate
the reflection and transmission coefficients, we set the left side input wave functions equation (\ref{Psi_L}) as
\begin{eqnarray}
f_{-2}^{k} &=&e^{-2ik}+r_{\mathrm{L}}e^{2ik},  \label{fL} \\
f_{-1}^{k} &=&e^{-ik}+r_{\mathrm{L}}e^{ik}, \\
f_{1}^{k} &=&t_{\mathrm{L}}e^{ik}, \\
f_{2}^{k} &=&t_{\mathrm{L}}e^{2ik},
\end{eqnarray}%
and the right side input wave functions equation (\ref{Psi_R})\ as
\begin{eqnarray}
f_{-2}^{k} &=&t_{\mathrm{R}}e^{2ik}, \\
f_{-1}^{k} &=&t_{\mathrm{R}}e^{ik}, \\
f_{1}^{k} &=&e^{-ik}+r_{\mathrm{R}}e^{ik}, \\
f_{2}^{k} &=&e^{-2ik}+r_{\mathrm{R}}e^{2ik},  \label{fR}
\end{eqnarray}

Substituting $f_{-2}^{k}$, $f_{-1}^{k}$, $f_{1}^{k}$, $f_{2}^{k}$ of
equations (\ref{fL}-\ref{fR}) into equations (\ref{L1}-\ref{L4}) and
eliminating $f_{A}^{k}$, $f_{B}^{k}$, we get equations of $r_{\mathrm{L}}$, $%
t_{\mathrm{L}}$, $r_{\mathrm{R}}$, $t_{\mathrm{R}}$ for the axial $\mathcal{%
PT}$-symmetric rhombic ring configuration. Through directly algebraic
calculation and simplification, we obtain the reflection and transmission
coefficients $r_{\mathrm{L}}$, $t_{\mathrm{L}}$, $r_{\mathrm{R}}$, $t_{%
\mathrm{R}}$ as functions of $k,\Phi ,\gamma $ given in equations (\ref%
{axialPTrLR}-\ref{axialPTtR}). Using the same procedure, we get the
reflection and transmission coefficients for the reflection $\mathcal{PT}$-symmetric rhombic
ring configuration. After substituting $f_{-2}^{k}$, $f_{-1}^{k}$, $%
f_{1}^{k} $, $f_{2}^{k}$ of equations (\ref{fL}-\ref{fR}) into equations (%
\ref{R1}-\ref{R4}) and eliminating $f_{A}^{k}$, $f_{B}^{k}$, we get
equations of $r_{\mathrm{L}}$, $t_{\mathrm{L}}$, $r_{\mathrm{R}}$, $t_{%
\mathrm{R}}$. Through directly algebraic calculation and simplification, we
obtain the reflection and transmission coefficients $r_{\mathrm{L}}$, $t_{%
\mathrm{L}}$, $r_{\mathrm{R}}$, $t_{\mathrm{R}}$ as functions of $k,\Phi
,\gamma $ given in equations (\ref{reflectionPTrL}-\ref{reflectionPTtLR}).
\newline

\noindent \textbf{Scattering states at the spectral singularities. }The
scattering coefficients diverge at the spectral singularities, to calculate
the scattering states, we have the wave functions of equation (\ref{Psi_L})
replaced by $f_{j<0}^{k}=A_{k}e^{ikj}+B_{k}e^{-ikj}$, $%
f_{j>0}^{k}=C_{k}e^{ikj}+D_{k}e^{-ikj}$. Substituting $f_{-2}^{k}$, $%
f_{-1}^{k}$, $f_{1}^{k}$, $f_{2}^{k}$ into equations (\ref{L1}-\ref{L4}) of
the axial $\mathcal{PT}$-symmetric rhombic ring configuration, we obtain the
coefficients satisfying $A_{k}=\mp iD_{k},B=\pm iC_{k}$ at the spectral
singularities that i) $k=\pi /2$,$\ \gamma =\pm 2\sin \left( \Phi /2\right)
\neq 0$; and ii)$\ \Phi =\pi $, $\gamma =\pm 2\sin k\neq 0$. These indicate
the scattering states are a self-sustained emission and a reflectionless absorption.
\section*{Acknowledgements}
This work is partly supported by the National Basic Research Program (973
Program) of China (Grant No. 2012CB921900). L.J. also appreciates the support of Nankai University Baiqing Plan foundation (Grant No. ZB15006104). X.Z.Z. also appreciates the support of National Natural Science Foundation of China (Grant No. 11505126) and PhD research startup foundation of Tianjin Normal University (Grant No. 52XB1415). Z.S. also appreciates the support of CNSF (Grant No. 11374163).

\section*{Author contributions statement}
L.J. conceived the idea, carried out the study. Z.S. supervised the project. L.J. wrote the manuscript with helpful suggestions from X.Z.Z., G.Z., and Z.S. All authors discussed the results and reviewed the manuscript. Correspondence and requests for materials should be addressed to L. Jin.

\newpage
\section*{Supplementary Information}

\begin{figure*}[tbh]
\includegraphics[ bb=0 0 525 480, width=8 cm, clip]{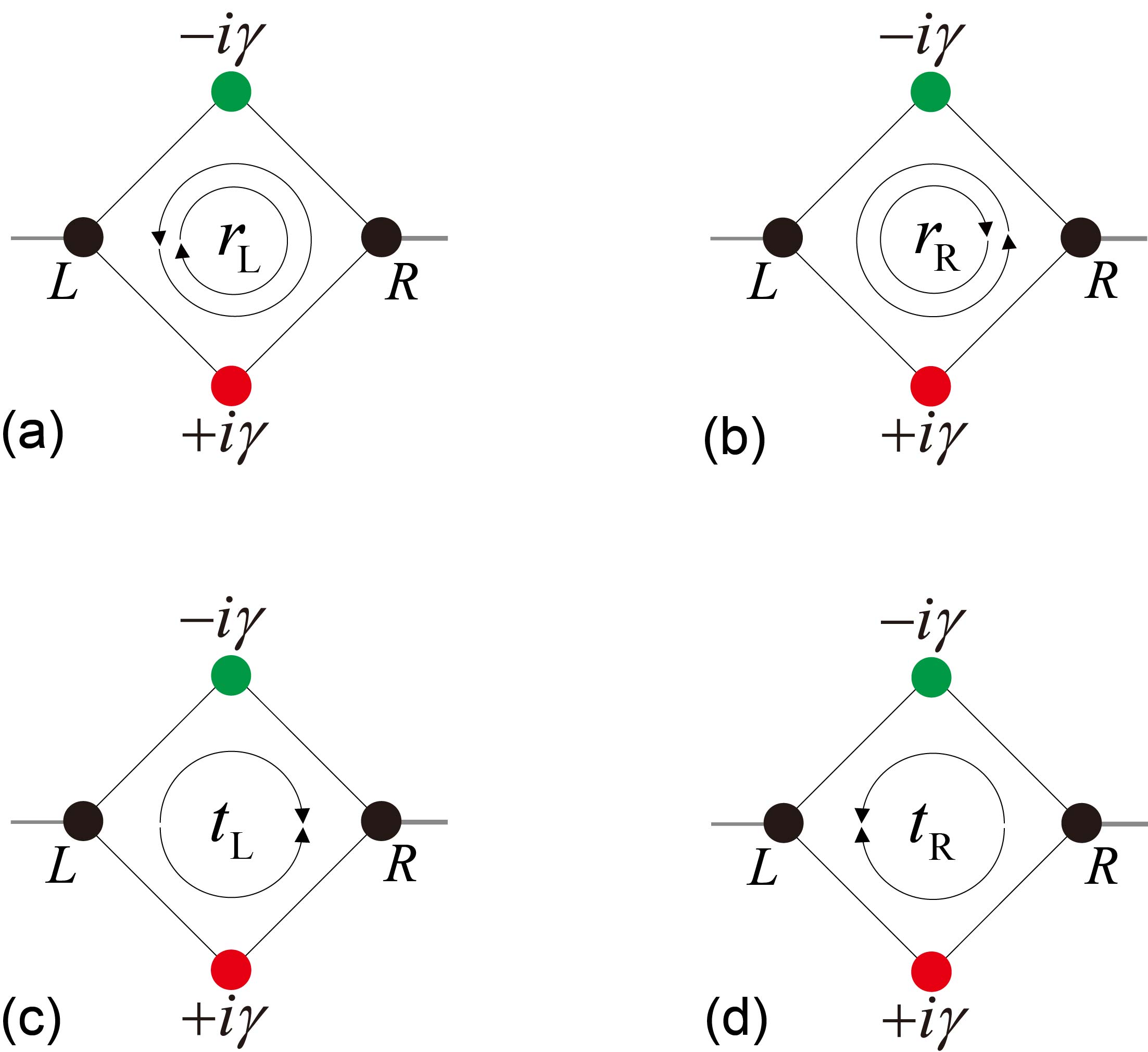}
\caption{\textbf{Photon pathways of axial $\mathcal{PT}$-symmetric rhombic
ring.} The arrows show the pathways in clockwise/counterclockwise direction
for (a) $r_{\mathrm{L}}$, (b) $r_{\mathrm{R}}$, (c) $t_{\mathrm{L}}$, (d) $%
t_{\mathrm{R}}$.}
\label{figS1}
\end{figure*}
\begin{figure*}[tbh]
\includegraphics[ bb=0 0 525 480, width=8 cm, clip]{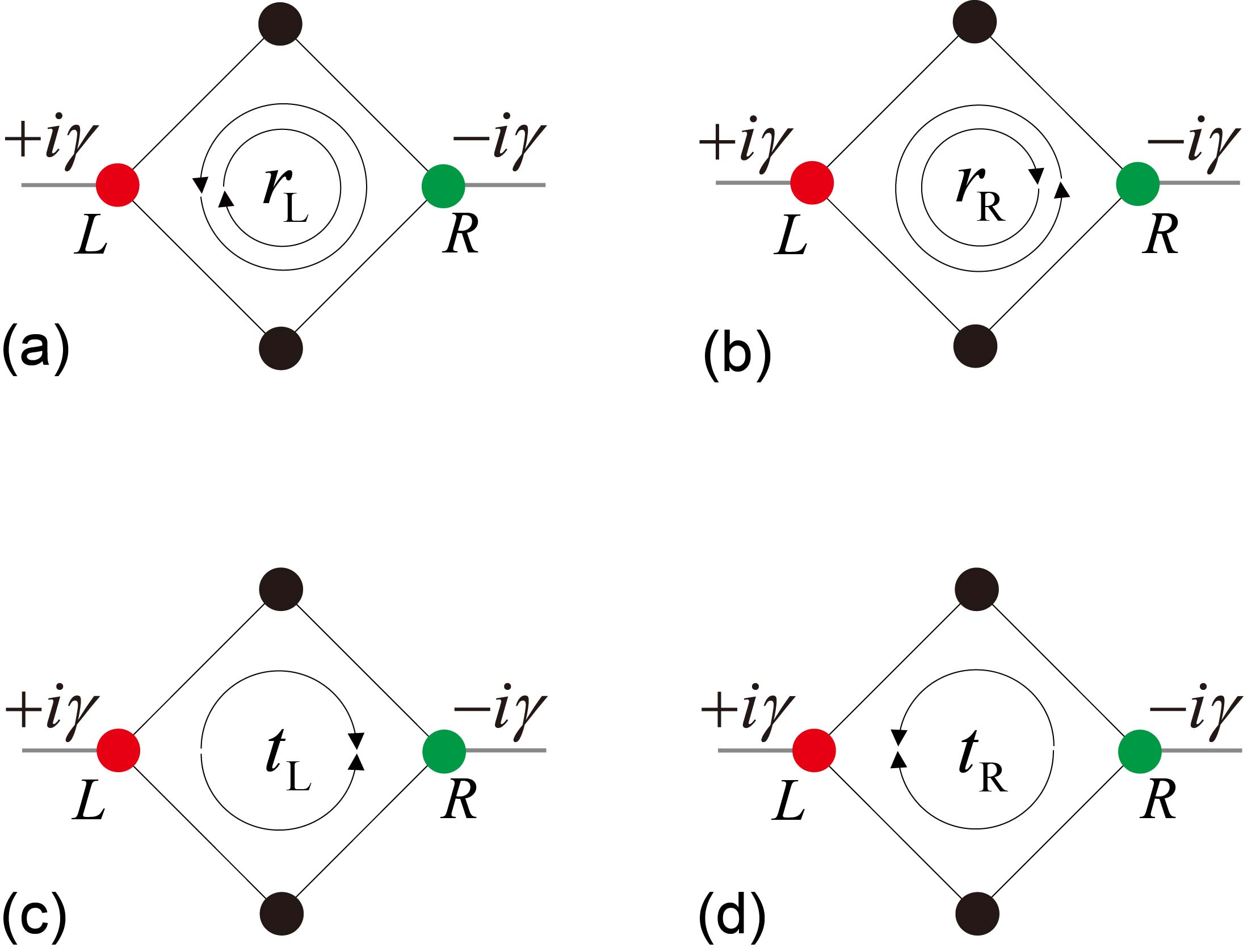}
\caption{\textbf{Photons pathways of reflection $\mathcal{PT}$-symmetric
rhombic ring.} The arrows show the pathways in clockwise/counterclockwise
direction for (a) $r_{\mathrm{L}}$, (b) $r_{\mathrm{R}}$, (c) $t_{\mathrm{L}}
$, (d) $t_{\mathrm{R}}$.}
\label{figS2}
\end{figure*}
\newpage In the Supplementary Information, we show the reciprocal reflection
(transmission) for axial (reflection) $\mathcal{PT}$ symmetry in details. The input and output leads are
described by two uniform semi-infinite tight-binding chains connected to the scattering centre. The wave functions for the left and right lead input with wave vector $k$ are $\psi _{\mathrm{L}%
}^{k}$ and $\psi _{\mathrm{R}}^{k}$, in form of
\begin{eqnarray}
\psi _{\mathrm{L}}^{k}\left( j\right) &=&\left\{
\begin{array}{c}
e^{ikj}+r_{\mathrm{L}}e^{-ikj},j<0 \\
t_{\mathrm{L}}e^{ikj},j>0%
\end{array}%
\right. ,  \label{Psi_L} \\
\psi _{\mathrm{R}}^{k}\left( j\right) &=&\left\{
\begin{array}{c}
t_{\mathrm{R}}e^{-ikj},j<0 \\
e^{-ikj}+r_{\mathrm{R}}e^{ikj},j>0%
\end{array}%
\right. .  \label{Psi_R}
\end{eqnarray}%
where $r_{\mathrm{L,R}}$ and $t_{\mathrm{L,R}}$ represent corresponding
reflection and transmission coefficients of wave with vector $k$.

\section*{Supplementary Note 1. Reciprocal reflection under axial $\mathcal{%
PT}$\ symmetry}

In this situation (Fig.~\ref{fig1}a), the whole scattering system is axial $\mathcal{%
PT}$-symmetric with respect to the leads, where $\left( \mathcal{PT}\right)
H_{\mathrm{L}}\left( \mathcal{PT}\right) ^{-1}$ $=H_{\mathrm{L}}$, $\left(
\mathcal{PT}\right) H_{\mathrm{R}}\left( \mathcal{PT}\right) ^{-1}$ $=H_{%
\mathrm{R}}$, and $\left( \mathcal{PT}\right) H_{\mathrm{c}}\left( \mathcal{%
PT}\right) ^{-1}$ $=H_{\mathrm{c}}$. The $\mathcal{PT}$ symmetry of the
scattering system results in symmetric relations on its wave functions. In
order to reveal the symmetry properties hidden in the scattering wave
functions, we act the $\mathcal{PT}$ operator on the wave function of left
lead input $\psi _{\mathrm{L}}^{k}\left( j\right) $, and get
\begin{equation}
\mathcal{PT}\psi _{\mathrm{L}}^{k}\left( j\right) =\left\{
\begin{array}{c}
e^{-ikj}+r_{\mathrm{L}}^{\ast }e^{ikj},j<0 \\
t_{\mathrm{L}}^{\ast }e^{-ikj},j>0%
\end{array}%
\right. ,
\end{equation}%
we act $\mathcal{PT}$ operator on the wave function of right lead input $\psi _{%
\mathrm{R}}^{k}\left( j\right) $, and get
\begin{equation}
\mathcal{PT}\psi _{\mathrm{R}}^{k}\left( j\right) =\left\{
\begin{array}{c}
t_{\mathrm{R}}^{\ast }e^{ikj},j<0 \\
e^{ikj}+r_{\mathrm{R}}^{\ast }e^{-ikj},j>0%
\end{array}%
\right. .
\end{equation}%
The real energy scattering state is $\mathcal{PT}$-symmetric due to the $%
\mathcal{PT}$ symmetry of the whole scattering system. Moreover, the two
series of wave functions (the eigenstates before and after acting the $%
\mathcal{PT}$ operator) are both eigenstates of the system with eigenvalue $%
E_{k}=-2\cos k$. Therefore, they must be in accords with each other, i.e.,
we can use $\psi _{\mathrm{L}}^{k}\left( j\right) $ and $\psi _{\mathrm{R}%
}^{k}\left( j\right) $ to compose $\mathcal{PT}\psi _{\mathrm{L}}^{k}\left(
j\right) $ and $\mathcal{PT}\psi _{\mathrm{R}}^{k}\left( j\right) $, because
the left side input and right side input scattering states are degenerate.

Composing $\mathcal{PT}\psi _{\mathrm{L}}^{k}\left( j\right) $ via $\psi _{%
\mathrm{L}}^{k}\left( j\right) $ and $\psi _{\mathrm{R}}^{k}\left( j\right) $
of equations (\ref{Psi_L}, \ref{Psi_R}) by eliminating $e^{ikj}$ in $j>0$
region and comparing the coeffients of $e^{\pm ikj}$ in the result with $%
\mathcal{PT}\psi _{\mathrm{L}}^{k}\left( j\right) $, we obtain%
\begin{eqnarray}
t_{\mathrm{L}}^{\ast }\left( t_{\mathrm{L}}t_{\mathrm{R}}-r_{\mathrm{L}}r_{%
\mathrm{R}}\right) &=&t_{\mathrm{L}},  \label{relation_axial3} \\
-r_{\mathrm{R}}t_{\mathrm{L}}^{\ast } &=&r_{\mathrm{L}}^{\ast }t_{\mathrm{L}%
},  \label{relation_axial4}
\end{eqnarray}%
Composing $\mathcal{PT}\psi _{\mathrm{R}}^{k}\left( j\right) $ via $\psi _{%
\mathrm{L}}^{k}\left( j\right) $ and $\psi _{\mathrm{R}}^{k}\left( j\right) $
of equations (\ref{Psi_L}, \ref{Psi_R}) by eliminating $e^{-ikj}$ in $j<0$
region and comparing the coeffients of $e^{\pm ikj}$ in the result with $%
\mathcal{PT}\psi _{\mathrm{R}}^{k}\left( j\right) $, we obtain%
\begin{eqnarray}
t_{\mathrm{R}}^{\ast }\left( t_{\mathrm{L}}t_{\mathrm{R}}-r_{\mathrm{L}}r_{%
\mathrm{R}}\right) &=&t_{\mathrm{R}},  \label{relation_axial1} \\
-r_{\mathrm{L}}t_{\mathrm{R}}^{\ast } &=&r_{\mathrm{R}}^{\ast }t_{\mathrm{R}%
},  \label{relation_axial2}
\end{eqnarray}%
we simplify the relations in equations (\ref{relation_axial3}, \ref%
{relation_axial4}), and get%
\begin{eqnarray}
|r_{\mathrm{L}}|^{2}+t_{\mathrm{L}}^{\ast }t_{\mathrm{R}} &=&1,
\label{relation_axial5} \\
r_{\mathrm{L}}^{\ast }t_{\mathrm{L}}+r_{\mathrm{R}}t_{\mathrm{L}}^{\ast }
&=&0,  \label{relation_axial6}
\end{eqnarray}%
we simplify the relations in equations (\ref{relation_axial1}, \ref%
{relation_axial2}), and get%
\begin{eqnarray}
|r_{\mathrm{R}}|^{2}+t_{\mathrm{L}}t_{\mathrm{R}}^{\ast } &=&1,
\label{relation_axial7} \\
r_{\mathrm{R}}^{\ast }t_{\mathrm{R}}+r_{\mathrm{L}}t_{\mathrm{R}}^{\ast }
&=&0,  \label{relation_axial8}
\end{eqnarray}%
From equations (\ref{relation_axial5}, \ref{relation_axial7}), we obtain the
reciprocal reflection for axial $\mathcal{PT}$ symmetry.%
\begin{equation}
|r_{\mathrm{L}}|^{2}=|r_{\mathrm{R}}|^{2}.
\end{equation}%
Moreover, considering the wave vector $k$ and $-k$, we obtain $r_{\mathrm{L}%
}(-k)=r_{\mathrm{L}}^{\ast }\left( k\right) $, $t_{\mathrm{L}}(-k)=t_{%
\mathrm{L}}^{\ast }(k)$ by comparing $\mathcal{PT}\psi _{\mathrm{L}%
}^{k}\left( j\right) $ and $\psi _{\mathrm{L}}^{-k}\left( j\right) $; and we
obtain $r_{\mathrm{R}}(-k)=r_{\mathrm{R}}^{\ast }\left( k\right) $, $t_{%
\mathrm{R}}(-k)=t_{\mathrm{R}}^{\ast }(k)$ by comparing $\mathcal{PT}\psi _{%
\mathrm{R}}^{k}\left( j\right) $ and $\psi _{\mathrm{R}}^{-k}\left( j\right)
$.

\section*{Supplementary Note 2. Reciprocal transmission under reflection $%
\mathcal{PT}$ symmetry}

In this situation (Fig.~\ref{fig1}b), the whole scattering system is reflection $%
\mathcal{PT}$-symmetric with respect to the leads, where $\left( \mathcal{PT}%
\right) H_{\mathrm{L}}\left( \mathcal{PT}\right) ^{-1}$ $=H_{\mathrm{R}}$, $%
\left( \mathcal{PT}\right) H_{\mathrm{R}}\left( \mathcal{PT}\right) ^{-1}$ $%
=H_{\mathrm{L}}$, and $\left( \mathcal{PT}\right) H_{\mathrm{c}}\left(
\mathcal{PT}\right) ^{-1}$ $=H_{\mathrm{c}}$. The $\mathcal{PT}$ symmetry of
the scattering system results in symmetric relations on its wave functions.
In order to reveal the symmetry properties hidden in the scattering wave
function, we act the $\mathcal{PT}$ operator on the wave function of the
left lead input $\psi _{\mathrm{L}}^{k}\left( j\right) $ similarly as the
previous analysis, and get%
\begin{equation}
\mathcal{PT}\psi _{\mathrm{L}}^{k}\left( j\right) =\left\{
\begin{array}{c}
t_{\mathrm{L}}^{\ast }e^{ikj},j<0 \\
e^{ikj}+r_{\mathrm{L}}^{\ast }e^{-ikj},j>0%
\end{array}%
\right. ,  \label{PT_PsiL_PT}
\end{equation}%
we act the $\mathcal{PT}$ operator on the wave function of right lead input $%
\psi _{\mathrm{R}}^{k}\left( j\right) $, and get%
\begin{equation}
\mathcal{PT}\psi _{\mathrm{R}}^{k}\left( j\right) =\left\{
\begin{array}{c}
e^{-ikj}+r_{\mathrm{R}}^{\ast }e^{ikj},j<0 \\
t_{\mathrm{R}}^{\ast }e^{-ikj},j>0%
\end{array}%
\right. .  \label{PT_PsiR_PT}
\end{equation}%
We take $\psi _{\mathrm{L}}^{k}\left( j\right) $ and $\psi _{\mathrm{R}%
}^{k}\left( j\right) $ to compose $\mathcal{PT}\psi _{\mathrm{L}}^{k}\left(
j\right) $ and $\mathcal{PT}\psi _{\mathrm{R}}^{k}\left( j\right) $.
Composing $\psi _{\mathrm{L}}^{k}\left( j\right) $ and $\psi _{\mathrm{R}%
}^{k}\left( j\right) $ of equations (\ref{Psi_L}, \ref{Psi_R}) in $j<0$
region by eliminating $e^{-ikj}$ and comparing the coeffients of $e^{\pm ikj}
$ in the result with $\mathcal{PT}\psi _{\mathrm{L}}^{k}\left( j\right) $,
we obtain%
\begin{eqnarray}
t_{\mathrm{L}}^{\ast }\left( t_{\mathrm{L}}t_{\mathrm{R}}-r_{\mathrm{L}}r_{%
\mathrm{R}}\right)  &=&t_{\mathrm{R}},  \label{relation_reflection1} \\
-t_{\mathrm{L}}^{\ast }r_{\mathrm{L}} &=&r_{\mathrm{L}}^{\ast }t_{\mathrm{R}%
},  \label{relation_reflection2}
\end{eqnarray}

Composing $\psi _{\mathrm{L}}^{k}\left( j\right) $ and $\psi _{\mathrm{R}%
}^{k}\left( j\right) $ of equations (\ref{Psi_L}, \ref{Psi_R}) in $j>0$
region by eliminating $e^{ikj}$ and comparing the coeffients of $e^{\pm ikj}$
in the result with $\mathcal{PT}\psi _{\mathrm{R}}^{k}\left( j\right) $, we
obtain%
\begin{eqnarray}
t_{\mathrm{R}}^{\ast }\left( t_{\mathrm{L}}t_{\mathrm{R}}-r_{\mathrm{L}}r_{%
\mathrm{R}}\right) &=&t_{\mathrm{L}},  \label{relation_reflection3} \\
-r_{\mathrm{R}}t_{\mathrm{R}}^{\ast } &=&r_{\mathrm{R}}^{\ast }t_{\mathrm{L}%
},  \label{relation_reflection4}
\end{eqnarray}%
we simplify the relations in equations (\ref{relation_reflection1}, \ref%
{relation_reflection2}), and get%
\begin{eqnarray}
|t_{\mathrm{L}}|^{2}+r_{\mathrm{L}}^{\ast }r_{\mathrm{R}} &=&1,
\label{relation_reflection5} \\
r_{\mathrm{L}}t_{\mathrm{L}}^{\ast }+r_{\mathrm{L}}^{\ast }t_{\mathrm{R}}
&=&0,  \label{relation_reflection6}
\end{eqnarray}%
we simplify the relations in equations (\ref{relation_reflection3}, \ref%
{relation_reflection4}), and get%
\begin{eqnarray}
|t_{\mathrm{R}}|^{2}+r_{\mathrm{L}}r_{\mathrm{R}}^{\ast } &=&1,
\label{relation_reflection7} \\
r_{\mathrm{R}}^{\ast }t_{\mathrm{L}}+r_{\mathrm{R}}t_{\mathrm{R}}^{\ast }
&=&0,  \label{relation_reflection8}
\end{eqnarray}%
From equations (\ref{relation_reflection5}, \ref{relation_reflection7}), we
obtain the reciprocal transmission for reflection $\mathcal{PT}$ symmetry.%
\begin{equation}
|t_{\mathrm{L}}|^{2}=|t_{\mathrm{R}}|^{2}.
\end{equation}%
Moreover, considering the wave vector $k$ and $-k$, we obtain $r_{\mathrm{L}%
}(-k)=r_{\mathrm{R}}^{\ast }\left( k\right) $, $t_{\mathrm{L}}(-k)=t_{%
\mathrm{R}}^{\ast }(k)$ by comparing $\mathcal{PT}\psi _{\mathrm{R}%
}^{k}\left( j\right) $ and $\psi _{\mathrm{L}}^{-k}\left( j\right) $; and we
obtain $r_{\mathrm{R}}(-k)=r_{\mathrm{L}}^{\ast }\left( k\right) $, $t_{%
\mathrm{R}}(-k)=t_{\mathrm{L}}^{\ast }(k)$ by comparing $\mathcal{PT}\psi _{%
\mathrm{L}}^{k}\left( j\right) $ and $\psi _{\mathrm{R}}^{-k}\left( j\right)
$.

\end{document}